\documentclass[11pt]{article}

\usepackage{amsmath}
\usepackage{amssymb}
\usepackage{bm}
\usepackage{color}
\usepackage{units}
\usepackage{graphicx}
\usepackage[letterpaper, margin=1in]{geometry}
\usepackage{authblk}
\usepackage{booktabs}
\usepackage{multirow}
\usepackage{afterpage}
\usepackage{filecontents}
\usepackage[version=3]{mhchem} 
\usepackage{textcomp}
\usepackage{gensymb}
\usepackage[normalem]{ulem}
\usepackage{mhchem}
\usepackage[symbol]{footmisc}
\usepackage[labelfont=bf]{caption}
\usepackage{xcolor, soul} 
\DeclareMathOperator{\Tr}{Tr}

\newcommand{\ve}[1]{\mathbf{#1}}

\definecolor{mygreen}{RGB}{0, 120, 0}

\newcommand{\beginsupplement}{%
        \setcounter{table}{0}
        \renewcommand{\thetable}{S\arabic{table}}%
        \setcounter{figure}{0}
        \renewcommand{\thefigure}{S\arabic{figure}}%
     }

\title{Swelling cholesteric liquid crystal shells to direct colloids at the interface}

\author[1]{Lisa Tran\footnote{Address correspondence to lt2727@columbia.edu.}}

\author[1]{Kyle J.M. Bishop} 

\affil{Department of Chemical Engineering, Columbia University, New York NY 10027, United States}

\date{}

\begin{document}

\maketitle

\begin{abstract}

Cholesteric liquid crystals can exhibit spatial patterns in molecular alignment at interfaces that can be exploited for particle assembly. These patterns emerge from the competition between bulk and surface energies, tunable with the system geometry. In this work, we use the osmotic swelling of cholesteric double emulsions to assemble colloidal particles through a pathway-dependent process. Particles can be repositioned from a surface-mediated to an elasticity-mediated state through dynamically thinning the cholesteric shell at a rate comparable to that of colloidal adsorption. By tuning the balance between surface and bulk energies with the system geometry, colloidal assemblies on the cholesteric interface can be molded by the underlying elastic field to form linear aggregates. The transition of adsorbed particles from surface regions with homeotropic anchoring to defect regions is accompanied by a reduction in particle mobility. The arrested assemblies subsequently map out and stabilize topological defects. These results demonstrate the kinetic arrest of interfacial particles within definable patterns by regulating the energetic frustration within cholesterics. This work highlights the importance of kinetic pathways for particle assembly in liquid crystals, of relevance to optical and energy applications.

\end{abstract}

\section*{Introduction}

Liquid crystal mesophases provide an attractive medium for assembling nanoparticle materials that can be tuned or reconfigured by external stimuli \cite{lc-app-1, gold-clc-1, gold-clc-2, gold-clc-3, gold-sm-yang, kumacheva-pnas, abbott-jdp-pnas}. Depending on their size, shape and surface chemistry, particles can disrupt the orientational order of liquid crystal molecules to create elastic distortions and topological defects, in which local nematic order is lost. Particles interact and assemble within liquid crystals so as to minimize the energetic penalties accompanying such distortions and defects. These energies are typically much larger than the thermal energy as characterized by the ratio, $K a / k_B T \gg 1$, where $K$ is the Frank elastic constant and $a$ is the particle radius (here, $K= 2\times10^{-11}$ N and $a= 100$ nm). As a result, particles may become kinetically trapped in local energy minima that depend on the processing history of the material.

Liquid crystals can also guide the assembly of particles at fluid interfaces due to surface patterns in molecular orientation. Previously, we showed that colloidal particles dispersed in aqueous solution can adsorb preferentially onto stripes of homeotropic (\textit{i.e.}, perpendicular) alignment that decorate droplets of a chiral liquid crystal \cite{lt-sciadv}. Such patterns arise due to competition between the preferred helical order in the bulk and the preferred homeotropic alignment at the interface induced by a molecular surfactant \cite{lt-prx}. Nanoparticles functionalized using the same surfactant associate favorably with those surface regions that have homeotropic anchoring (Fig.~\ref{ResultsIntro}A, left \& Fig.~\ref{ResultsIntro}E, i-ii). The resulting particle assemblies at liquid crystal interfaces are mediated by molecular surface forces and are therefore distinct from those mediated by elastic distortions and defects in the bulk \cite{lt-sciadv}. 

Here, we describe how nanoparticle assemblies at liquid crystal interfaces can be dynamically reconfigured from one state stabilized by surface forces to another stabilized by elastic distortions and defects in a pathway-dependent manner. We use microfluidics to prepare spherical shells of a cholesteric liquid crystal, which exhibits complex surface patterns due to the competing effects of geometrical confinement, surface anchoring, and bulk ordering. In contrast to similar patterns on droplets \cite{lt-sciadv}, those on shells can be dynamically tuned by varying the shell thickness through osmotic swelling \cite{afn-nematic, teresa-nematic-shells}. Colloidal particles introduced to the outer aqueous phase adsorb strongly at the liquid crystal interface but remain mobile and free from aggregation due to added surfactants. During swelling, the adsorbed particles are repositioned from surface patterns with homeotropic anchoring (Fig.~\ref{ResultsIntro}A, left) into linear arrays along newly-formed defect lines (Fig.~\ref{ResultsIntro}A, right). The resulting assemblies depend sensitively on their kinetic history and are characterized by a sharp reduction in particle mobility. The nanoparticles further influence the cholesteric patterns, stabilizing topological defects under out-of-equilibrium conditions. Together, our results show how liquid crystals can provide dynamic energy landscapes that guide transitions among different types of nanoparticle assemblies. The creation of such tunable structures with submicron features over large areas of chemically accessible interfaces may prove useful in the design of materials with coupled chemo-electro-optical responses \cite{lc-app-1,sensing}.

\section*{Results}

\subsection*{Templating particle adsorption with cholesterics}

\begin{figure*}
\centerline{\includegraphics[width=0.9\textwidth]{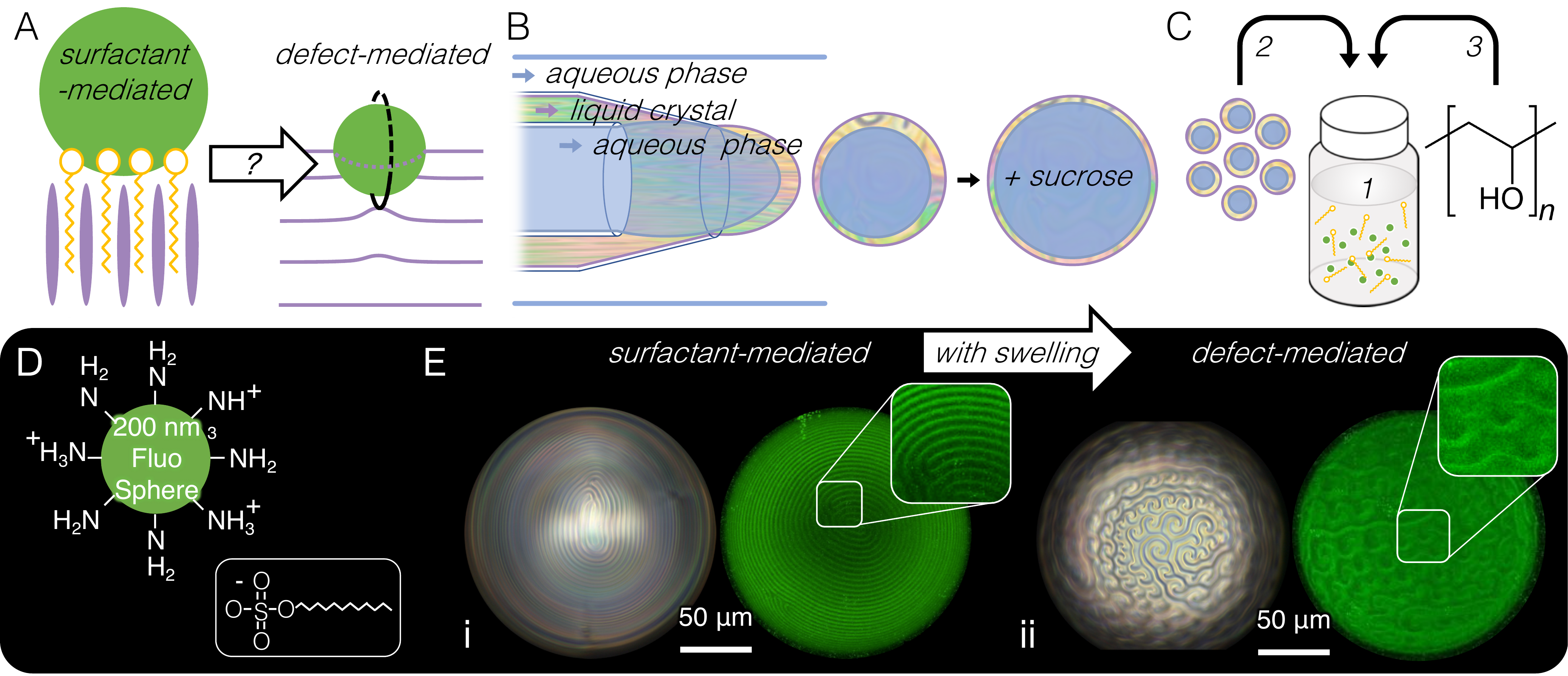}}
\caption{A) Illustrations of a physico-chemically favorable, surfactant-mediated particle state and an elastically-favorable, defect-mediated particle state (not drawn to scale). The left schematic represents a particle (green) with surfactants (yellow) that are ordered by the liquid crystal (purple). The right cartoon represents a particle (green) in a defect-bound state. Purple lines indicate the director field, while the particle-induced, half ``Saturn ring" defect at the liquid crystal interface is drawn in black. B) Sketch of the microfluidic device used to fabricate liquid crystal shells, along with osmotic swelling of shells using sucrose. C) Illustration of the step-wise addition of liquid crystal emulsions and PVA to the surfactant and nanoparticle mixture, which speeds up particle adsorption to the liquid crystal-water interface. D-E) Crossed-polarized images of cholesteric shells with nanoparticles at their interface (left) are juxtaposed against their fluorescence images (right). Amine-functionalized, 200 nm particles treated with 7-10 mM SDS (D) can create patterned crusts or segregated nanoparticle assemblies that are stable for months (E). Amines and SDS interact with PVA to moderate aggregation in the bulk and at the interface, finely templating particles (E). Particles typically conform to regions of homeotropic anchoring (E-i). However, when shells with adsorbed particles are osmotically swelled, the particles form dense assemblies along defect lines (E-ii). \label{ResultsIntro}}
\end{figure*}

We use a cholesteric liquid crystal (CLC) to generate a variety of patterns in molecular orientation at the water-CLC interface. The rod-like molecules of the CLC have an energetic tendency to align with one another and also to stack in a helical fashion, resulting in an overall twist of the director $\ve{n}$ along the pitch axis. This twist often leads to geometrical frustration. For instance, a surface imposing homeotropic anchoring opposes the natural winding of the cholesteric. The helical axis can reorient parallel to the surface to obtain some perpendicular alignment; however, this configuration results in periodic violations of the anchoring condition. For weak anchoring, the bulk twist dominates, and alternating stripes of perpendicular and tangential alignment are formed at the surface \cite{lt-prx}. By contrast, strong homeotropic anchoring acts to repel regions of planar alignment away from the surface, forming disclinations where the cholesteric twists rapidly  \cite{homeotdrop, zumerknot, dropskyrm}. The surface patterns that arise from this frustration between bulk twist and surface anchoring can be tuned by varying the thickness of the CLC phase \cite{lt-prx,alex-waltz}.

We use microfluidics to prepare water-in-CLC-in-water double emulsions, which provide free-standing CLC films of tunable thicknesses (Fig.~\ref{ResultsIntro}B; Methods). The cholesteric is composed of a nematic liquid crystal, 4-cyano-4'-pentylbiphenyl (5CB), mixed with a chiral dopant, (S)-4-cyano-4-(2-methylbutyl)biphenyl (CB15). The pitch $p$ is set by the dopant concentration \cite{cb15pitch}, which is fixed at 2.8\% wt such that $p\approx 5~\mu$m as verified by a Grandjean-Cano wedge cell \cite{GJWedge,Cwedge}. The inner and outer aqueous phases contain 1\% wt polyvinyl alcohol (PVA), which helps to stabilize the emulsions for many months when stored in a closed glass vial.  We add sucrose in different concentrations to the inner and outer aqueous phases to induce osmotic swelling of the CLC shells (Fig.~\ref{ResultsIntro}B)  \cite{hughes-osmswell,lee-osmswell,tll-waltz-patch}; the rate of swelling can be tuned by varying the concentration difference. As the inner aqueous volume increases, the liquid crystal volume remains constant, leading to a decrease in shell thickness over time. 

The addition of a molecular surfactant, sodium dodecylsulfate (SDS), induces adaptive, homeo-tropic anchoring conditions at the outer CLC-water interface \cite{tll-sds, lt-prx}. Using SDS concentrations of 7-10 mM, we observe the formation of periodic stripes within the liquid crystal with spacings and morphologies that depend on the shell thickness \cite{lt-prx, lt-sciadv}. The aliphatic tails of the adsorbed surfactants act to orient the liquid crystal molecules perpendicular to the interface. At the same time, the surfactant adsorbs preferentially onto surface regions with such homeotropic alignment \cite{abblipid,abblipid-2}. The interplay between surfactant adsorption and liquid crystal alignment leads to adaptive anchoring, whereby the surfactant and the liquid crystal mutually influence one another. This behavior differs from homeotropic anchoring at solid surfaces, where the anchoring strength is fixed to a uniform value across the surface.

\begin{figure*}[p]
\centerline{\includegraphics[width=0.48\textwidth]{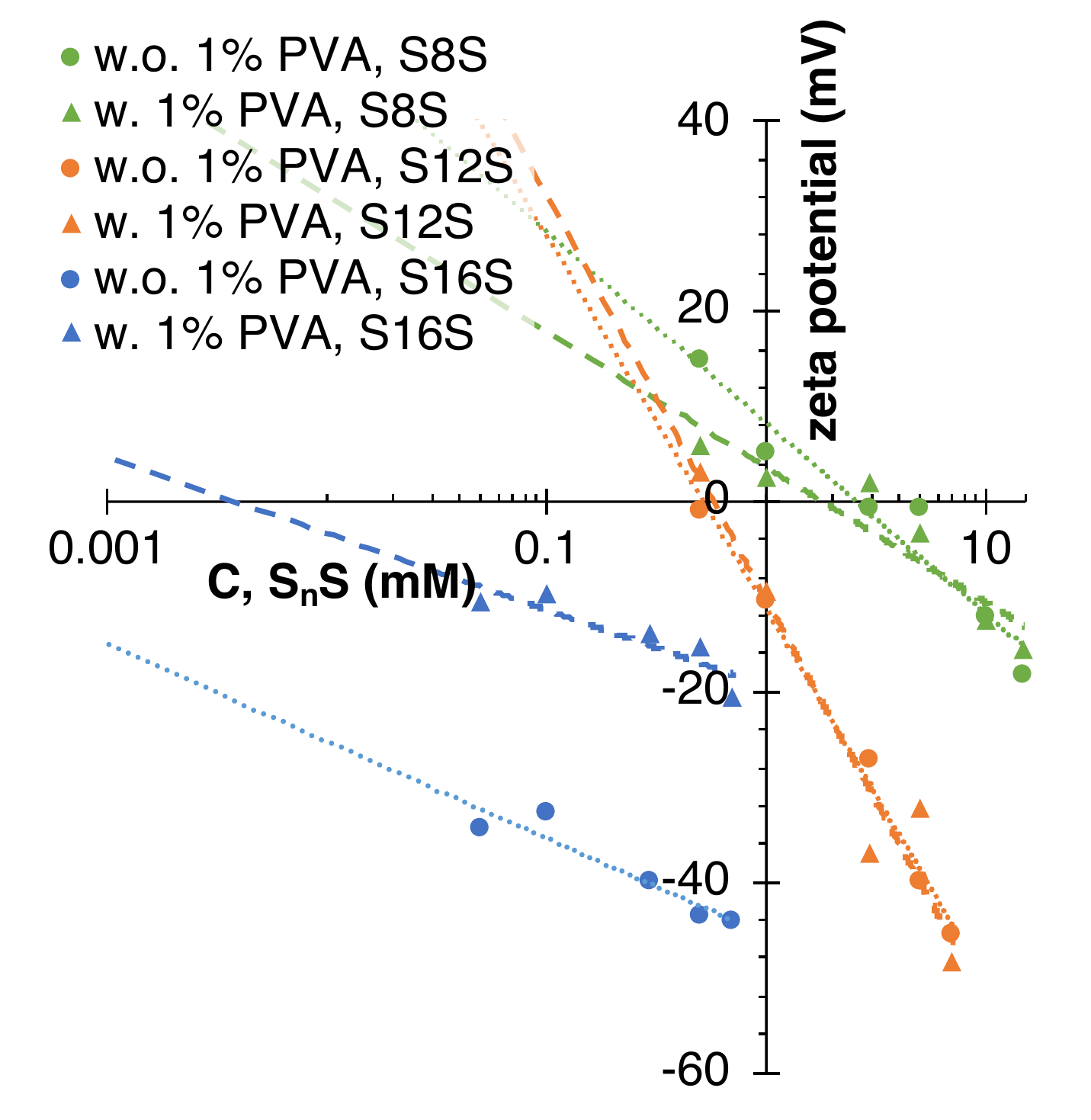}}
\caption{The zeta potential of 10 nm, Ludox-CL particles at 2\% wt is plotted against varying concentrations of sodium alkyl sulfate, S$_{n}$S, with $n$ representing the hydrocarbon tail length; $n$ = 8, 12, and 16 are measured, with and without 1\% wt PVA in the solution. Logarithmic fits of the data are plotted to guide the eye. PVA tends to suppress the zeta potential at low S$_{n}$S concentrations across all samples. For $n$ = 8 (sodium octyl sulfate, SOS) and 12 (sodium dodecylsulfate, SDS), the zeta potential of particles converge with increasing surfactant concentration. However, for $n$ = 16 (sodium hexadecyl sulfate, SHS), the zeta potentials between samples with and without PVA diverge greatly, with PVA significantly reducing the measured charge. The data suggests that the PVA backbone associates with aliphatic tails. \label{Zeta}}
\end{figure*}

Nanoparticles with appropriate surface functionalization adsorb at the CLC-water interface to create assemblies templated by the liquid crystal patterns \cite{lt-sciadv, lt-langmuir}. We use 200 nm FluoSpheres---fluorescent polystyrene particles with amine-functionalized surfaces---dispersed in solutions of the same SDS surfactant that induces homeotropic anchoring. At neutral pH, the negatively charged head groups of the surfactant adsorb onto the positively charged amine groups on the particles to create hydrophobic surfaces that interact favorably with the CLC-water interface. Importantly, undesired aggregation of these hydrophobic particles in aqueous solution is mitigated by the presence of the PVA polymer, which interacts with the particle surface (Supplementary Fig.~\ref{Zeta}). Under these conditions, the particles adsorb preferentially onto regions of the cholesteric patterns with homeotropic alignment as evidenced by cross-polarized light and fluorescence microscopy (Fig.~\ref{ResultsIntro}E, i-ii).

\begin{figure}
\centerline{\includegraphics[width=0.55\textwidth]{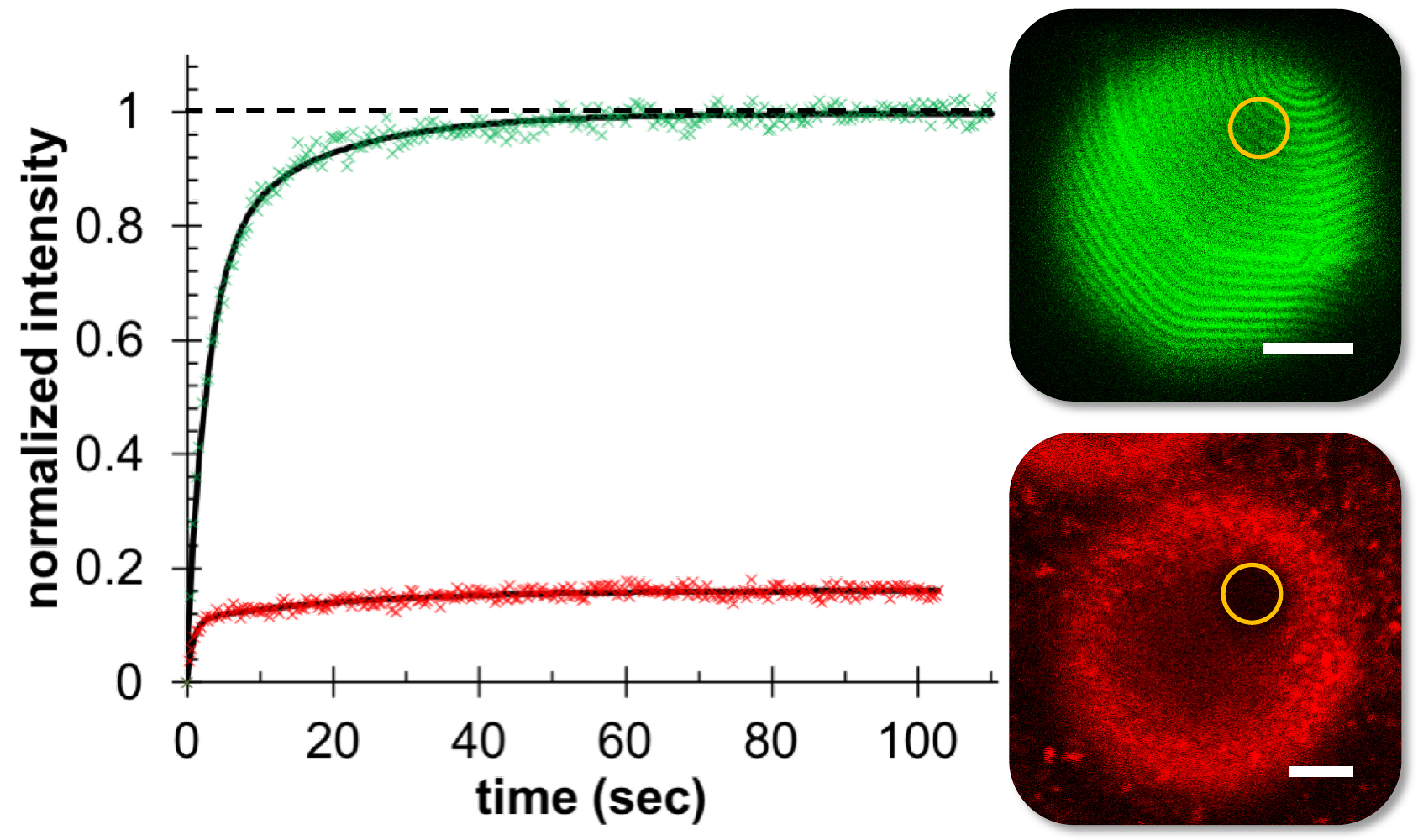}}
\caption{Fluorescence recovery after photobleaching (FRAP) measurements of nanoparticles at the cholesteric shell interface. Fluorescent particles, made hydrophobic by SDS (green) and dodecyltrimethylammonium bromide (DODAB) (red), are adsorbed onto shells (insets). Data points are plotted in color, and exponential fits of the data are plotted as solid black lines. Dashed black line marks 100\% fluorescence recovery. Scale bars are 25 $\mu$m. \label{FRAP}}
\end{figure}

\begin{figure*}[p]
\centerline{\includegraphics[width=0.9\textwidth]{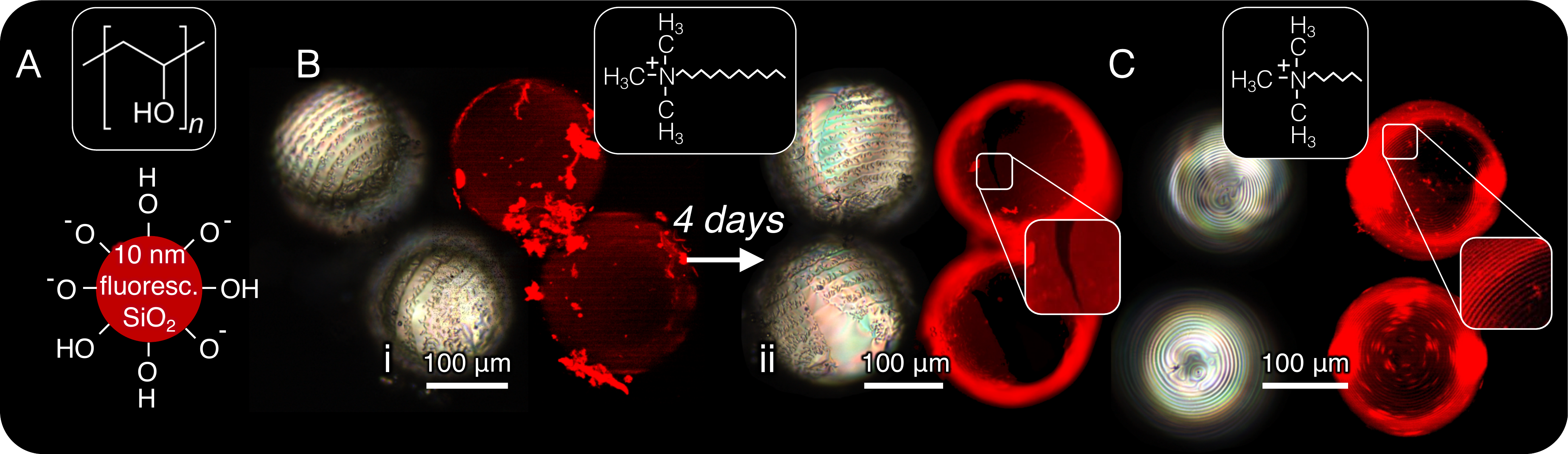}}
\caption{Crossed-polarized images of cholesteric shells (left) with nanoparticles at their interface are juxtaposed against their fluorescence images (right). 10 nm, TRITC-core, silica nanoparticles (A, bottom) are templated on cholesteric shells stabilized with PVA (A, top) using alkylammonium bromides (ATABs).  Nanoparticle crusts are created when particles are functionalized with dodecyltrimethylammonium bromide (DODAB) (B-i). A 1D periodic array of distorted disclination lines is formed, which is evidence of strong homeotropic anchoring, frustration of cholesteric pitch, and pinning of disclination lines. After 4 days, the shells crack open, revealing bare cholesteric interface with no stripes, signifying planar alignment (B-ii). When hexyltrimethylammonium bromide (HTAB) is used in place of DODAB, nanoparticles segregate into stripes along homeotropic anchoring (C). Assemblies formed using ATABs destabilize after a few days. \label{ATAB}}
\end{figure*}

The adsorbed particles remain highly mobile at the interface as evidenced by diffusivity measurements using fluorescence recovery after photobleaching (FRAP)  \cite{frap-1,frap-2,frap-3}.  The normalized fluorescence recovery for the FluoSpheres is plotted against time in Fig.~\ref{FRAP} (green markers).  The halftime of fluorescence recovery ranges from 1-10 seconds and reflects the speed at which unbleached particles diffuse into the 10 $\mu$m diameter bleached region (Fig.~\ref{FRAP}, insets). Using this data and the Soumpasis equation\cite{frap-1,frap-3}, the particle diffusion coefficient is estimated to be $\sim$10$^{-12}$ m$^{2}$/s , which is comparable to the Stokes-Einstein diffusivity of the particles in solution.  Moreover, for FRAP measurements performed within the same day of emulsion preparation, the fluorescence intensity is fully recovered after bleaching, suggesting that all particles remain mobile at the interface.  By contrast, experiments using a different system based on 10 nm silica particles functionalized with a cationic surfactant showed that only 20\% of those particles remain mobile at the interface (Fig.~\ref{FRAP}; red markers) \cite{frap-2}. The silica particles aggregate at the liquid crystal interface to form rigid crusts, which are unable to adapt to changes in the underlying surface patterns (Supplementary Fig.~\ref{ATAB}).

\subsection*{From adaptive anchoring to kinetic arrest}

\begin{figure*}
\centerline{\includegraphics[width=1\textwidth]{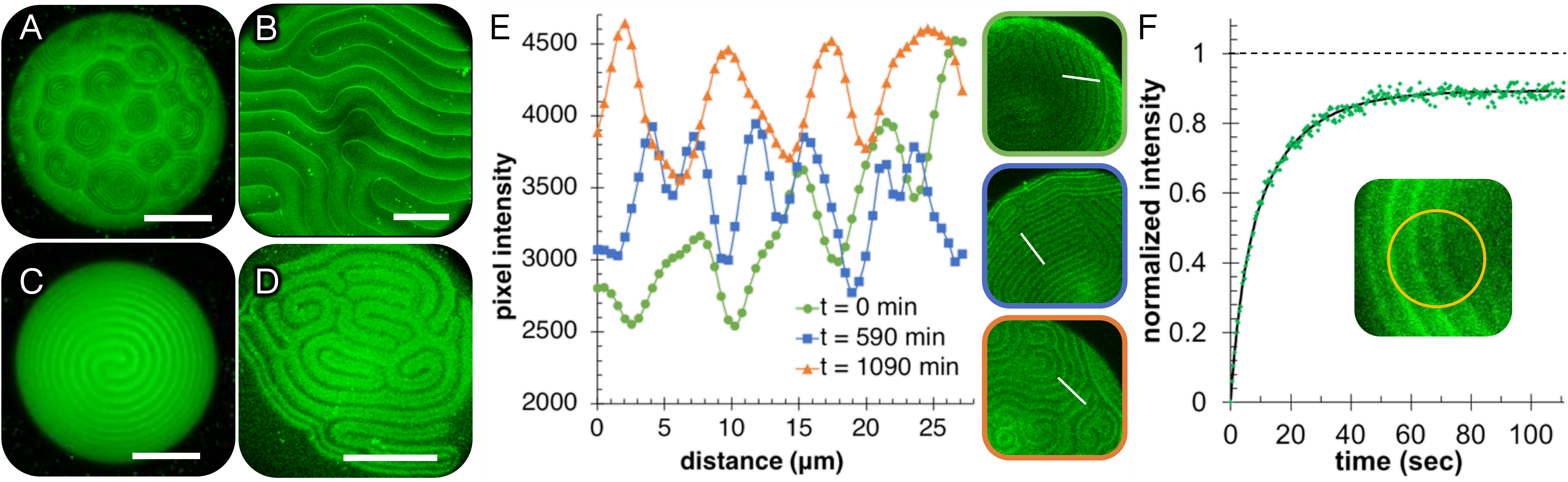}}
\caption{A-D) Osmotic swelling of cholesteric shells can shift adsorbed nanoparticles from regions of homeotropic anchoring to disclination lines as evidenced by fluorescence confocal micrographs. Scale bars are 50 $\mu$m. A-B) As shells are swelled with interfacially adsorbed particles in 7 mM SDS, the particle assemblies transition from following the anchoring (representative state in A) to following disclinations (representative state in B). C) For shells that are not swelled and have increased homeotropic alignment with 10 mM SDS, nanoparticles continue to locate along regions with homeotropic anchoring, despite the presence of disclinations. D) When shells are swollen before FluoSpheres are introduced, disclination lines are not formed, and particles follow cholesteric ``fingers''. E) Plots of pixel intensity over distance across stripes (white lines, 28 $\mu$m, insets) during swelling shows the transition of particles from homeotropic regions to defect lines. F) FRAP measurements of shells with nanoparticles assembled on disclinations (inset) reveal that the fluorescence intensity does not fully recover after bleaching within the yellow circle (diameter is 41 $\mu$m). \label{AnchtoDefect}}
\end{figure*}

\begin{figure*}
\centerline{\includegraphics[width=0.6\textwidth]{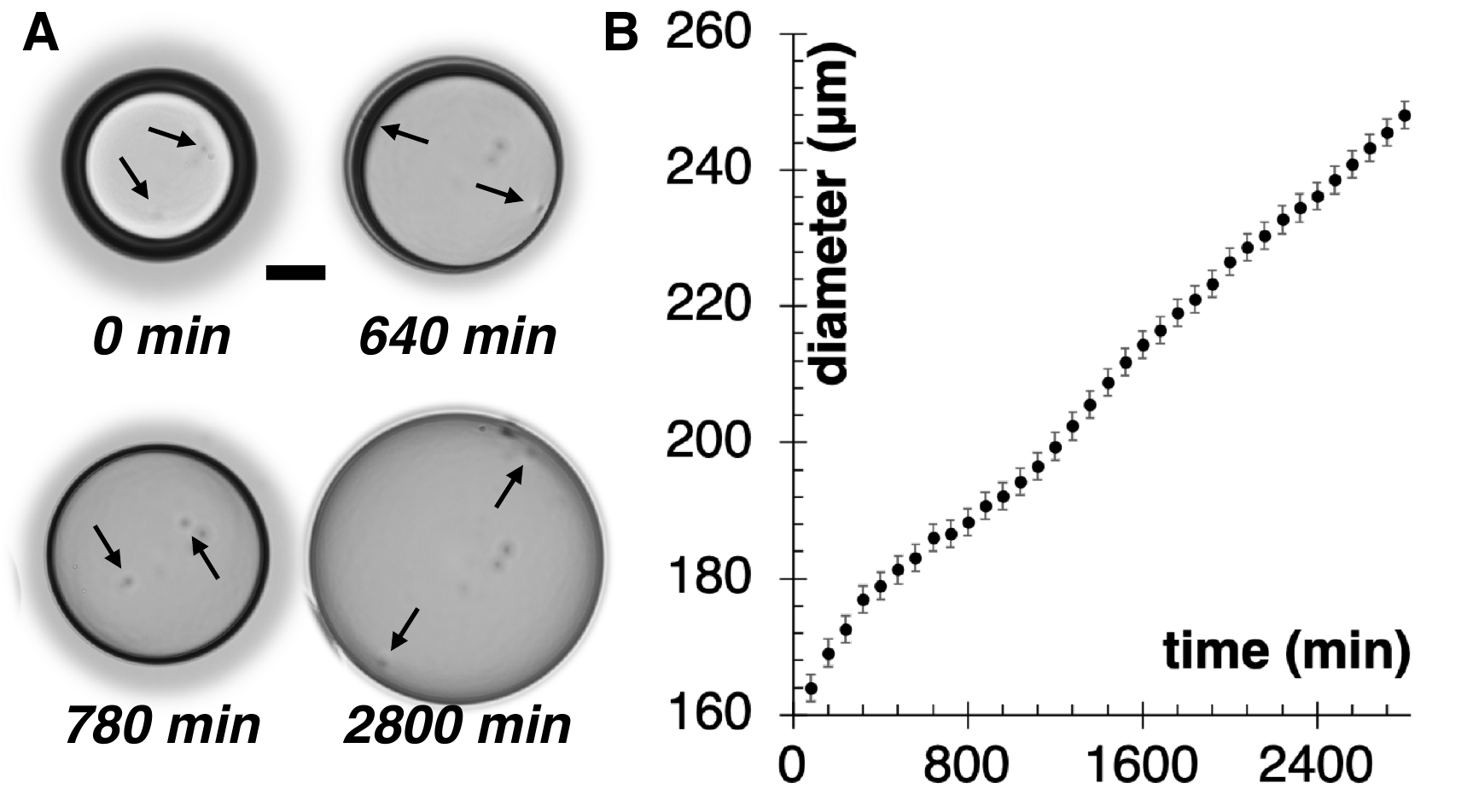}}
\caption{The rate of osmotic swelling for cholesteric shells is slower than particle diffusion. Scale bars are 50 $\mu$m. A) In i), cholesteric shells with only 1\% wt PVA in the aqueous phases are osmotically swollen. Arrows in optical micrographs point out two +1 defects reorienting during the swelling process. Optical micrographs reveal how the defects, and thereby the elastic field, change during swelling. B) Plots of the diameter versus time reveal that shells swell at a rate of $\sim$0.001 $\mu$m/sec, much slower than the nanoparticle rate of diffusion (1 $\mu$m$^2$/sec). \label{CholSwell}}
\end{figure*}

Osmotic swelling of the liquid crystal shells acts to reposition the adsorbed particles from regions of homeotropic anchoring (Fig.~\ref{AnchtoDefect}A) to newly-formed defect lines (Fig.~\ref{AnchtoDefect}B). CLC shells are prepared with 0.5 M sucrose in the inner aqueous phase and no sucrose in the outer phase, resulting in osmotic swelling at rates of $\sim$0.001 $\mu$m/s (Supplementary Fig.~\ref{CholSwell}). Addition of the FluoSphere-surfactant solution induces adaptive homeotropic anchoring at the liquid crystal interface, resulting in the formation of striped patterns as the particles adsorb from solution.  As discussed above, the particles are mobile on the interface and concentrate in regions of the stripe patterns with homeotropic alignment. Over time, the osmotic thinning of the shells and the adsorption of the particles act together to alter the free energy of the system, reducing the influence of the bulk cholesteric twist and increasing that of homeotropic anchoring. At some point in this kinetic process, regions of planar alignment at the interface are pushed into the bulk, forming disclinations. Fluorescence micrographs reveal that the adsorbed particles organize within linear chains along these defect lines (Fig.~\ref{AnchtoDefect}B). Supplementary Movie 1 illustrates this dynamic process whereby osmotic swelling, particle adsorption, and liquid crystal patterning guide the creation of particle assembles. 

The organization of particles along defect lines depends on the kinetic pathway by which the system is prepared.  When cholesteric emulsions are not swelled, particles adsorb onto regions of homeotropic alignment---even when disclinations are present. This behavior is demonstrated in Fig.~\ref{AnchtoDefect}C, where a shell in a solution of higher SDS concentration (10 mM) has defect lines with a low fluorescence intensity, indicating low particle density. Similarly, when particles are added \textit{after} swelling of the shells, disclinations are not present. Disclinations that are formed are annealed out of the system during osmotic swelling. Instead of along disclinations, the particle density is greatest along structures identifiable as cholesteric ``fingers'' (Fig.~\ref{AnchtoDefect}D) \cite{oswald2000static}, suggesting that the adsorbed particles play a role in nucleating disclinations. Only when shells are swelled in the presence of particles, do the resulting assemblies align along defect lines. Intriguingly, when shells are swelled more quickly using 0.8 M sucrose in the inner aqueous phase, the particles again follow cholesteric fingers, exhibiting patterns similar to those in Fig.~\ref{AnchtoDefect}D. These control experiments suggest that the kinetic processes of osmotic swelling and particle adsorption must proceed concurrently with the appropriate rates to achieve high particle densities along disclination lines, regardless of the shell's initial condition (Supplementary Movie 2).

Fluorescence confocal video microscopy reveals the dynamic reorganization of adsorbed particles from homeotropic stripes to defect lines (Supplementary Movie 3). In order to slow the evolution of the cholesteric patterns, we transfer shells from the particle dispersion after several minutes and place them in an otherwise identical solution without particles (10 mM SDS, 0.25 mM HCl, 1\% wt PVA). Fig.~\ref{AnchtoDefect}E shows the fluorescence intensity along scans perpendicular to the stripes (marked by a white line in insets) for multiple time steps during the swelling process.  At $t=0$ (green), adsorbed particles are located in homeotropic regions. From $t=0$ (green) to $t=590$ minutes (blue), particles begin to gather at the edges of homeotropic stripes, forming secondary peaks in the intensity (blue). From $t=590$ (blue) to $t=1090$ minutes (orange), the pattern evolves during swelling until only primary peaks along disclination lines remain.  The final particle assemblies are concentrated along defects (orange) located in regions that formerly exhibited planar alignment.

\begin{figure*}
\centerline{\includegraphics[width=0.48\textwidth]{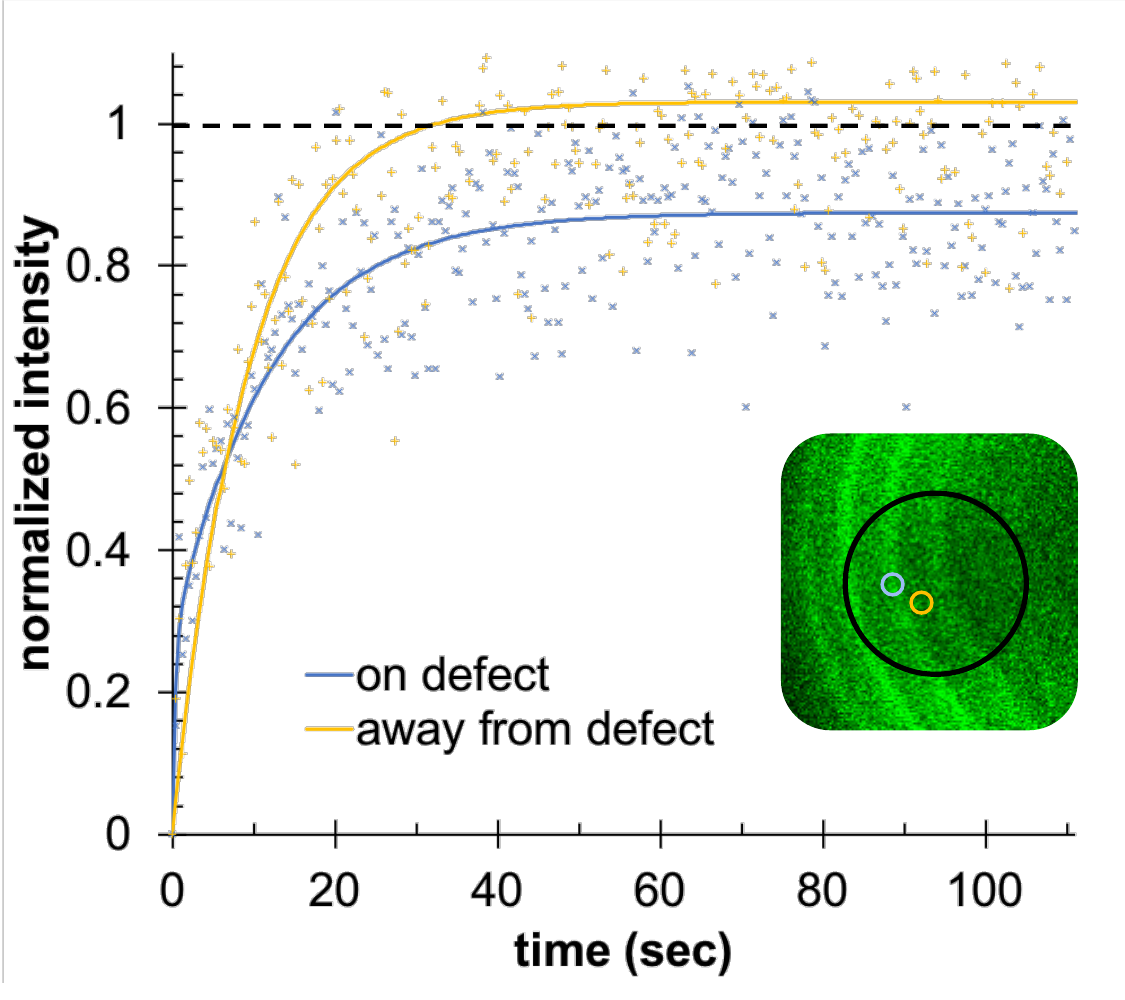}}
\caption{FRAP measurements of the system in Fig.~\ref{AnchtoDefect}E with nanoparticles assembled on disclinations (inset). Within the bleached region (black circle), the fluorescence recovery of nanoparticles along disclination lines versus along homeotropic anchoring differ from one another. The average percent of mobile species for nanoparticles along defect lines (blue) is $\sim$ 85\%.  The average percent of mobile species for nanoparticles along regions of homeotropic anchoring is $\sim$ 100\% (yellow). Due to the high scanning speed necessary for FRAP measurements, both signals are noisy with a standard deviation of $\sim$ 10\% for each. Despite noise due to the low resolution of regions examined, this difference in the percentage of mobile species points to distinct diffusive behaviors and provides support for our hypothesis of two nanoparticle populations at the interface. Diameters of regions (blue and yellow circles) measured within the bleaching radius in the inset is $\sim$ 4 $\mu$m. \label{FRAPDiffSec}}
\end{figure*}

FRAP measurements of particles assembled along disclination lines indicate a reduction in particle mobility (Fig.~\ref{AnchtoDefect}F).  The percentage of fluorescence recovery drops by $\sim$15\% as compared to particles adsorbed at homeotropic regions (\textit{cf.} Fig.~\ref{FRAP}), suggesting that some fraction of the particles are immobile. The remaining mobile particles (\textit{ca.} 85\% of the total) are characterized by a similar diffusivity to that quoted above.  Further analysis of the FRAP data suggests that the immobile particles are localized within the defect regions (Supplementary Fig.~\ref{FRAPDiffSec}).  Together, the observed dependence of the particle assemblies on their kinetic history and the reduction in particle mobility support a picture in which the adsorbed particles and the liquid crystal defects are guided in time to a deep energy well of mutual attraction. To further substantiate this picture, we turn now to simulations that describe the changing energy landscape of the liquid crystal phase.

\begin{figure}
\centerline{\includegraphics[width=0.55\textwidth]{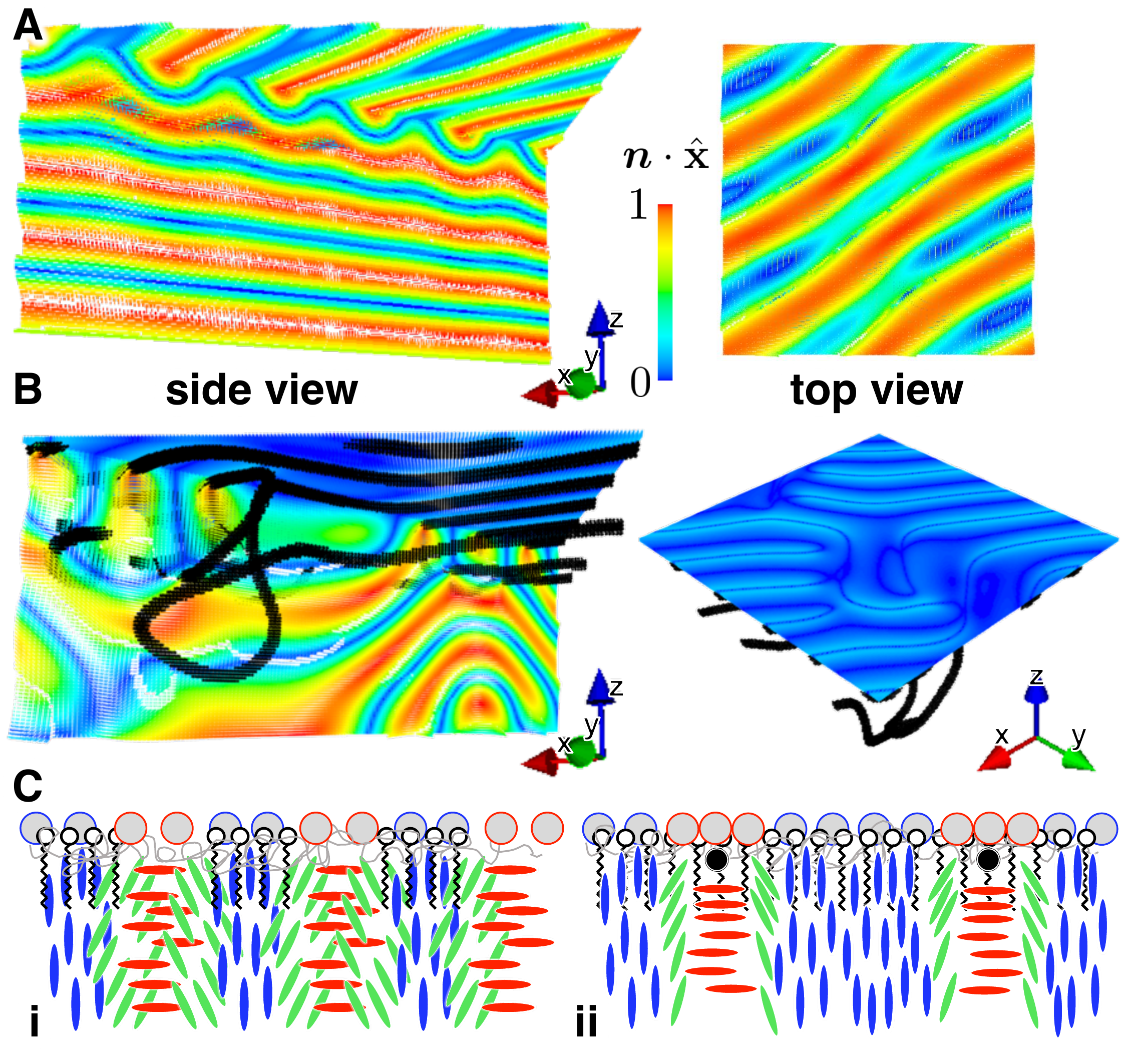}}
\caption{Landau-de Gennes simulations of cholesteric slabs with varying homeotropic anchoring strengths (weak for A, strong for B) at the top surface and planar anchoring at the bottom surface illustrate how director orientations (plotted with a color map of the horizontal component of the director $\mathbf{n} \cdot \mathbf{\hat{x}}$) change with increased anchoring strength. Defects are plotted in black. Schematic of possible nanoparticle reordering for a shift from weak to strong homeotropic anchoring is shown in C (not drawn to scale). Particles associated with homeotropic and planar anchoring are outlined in blue and red, respectively. \label{SimulationsMain}}
\end{figure}

Landau de-Gennes simulations of the CLC reproduce the transition from stripes to disclinations upon increasing the strength of homeotropic anchoring (Fig.~\ref{SimulationsMain}). As detailed in the Methods, we consider a rectangular slab (0.74 by 0.74 by 0.49 $\mu$m$^3$) of a CLC with a pitch of $p=0.25~\mu$m and periodic boundary conditions in the $x$ and $y$ directions. Strong planar anchoring is enforced at the bottom surface using the Rapini-Papoular potential with an anchoring strength of $W_1 = 4 \times 10^{-3}$ J/m$^2$. We vary the strength of homeotropic anchoring at the top surface from $W_0 = 8 \times 10^{-5}$ to $4 \times 10^{-3}$ J/m$^2$, which corresponds to a transition from weak to strong anchoring. For weak homeotropic anchoring (here, $K / p W_0 \sim 1$), stripes of parallel and perpendicular alignment form at the top surface with a characteristic spacing comparable to the pitch (Fig.~\ref{SimulationsMain}A).  For strong anchoring (here, $K/\xi W_0\sim 1$ where $\xi=6.6$ nm is the characteristic defect size), the stripes of planar alignment are repelled from the surface into the bulk, forming defect lines in which local nematic order is lost (Fig.~\ref{SimulationsMain}B). The simulations suggest that disclination lines form near the interface where the energetic contributions due to surface anchoring are comparable to those due to elastic distortions---that is, $\xi^2 W_0\sim \xi K\sim 100 k_B T$.

We hypothesize that adsorbed particles initially located in regions of planar alignment become kinetically arrested along defect lines as the liquid crystal patterns evolve (Fig.~\ref{SimulationsMain}C).  While the adsorbed particles are enriched in regions of homeotropic anchoring, they remain distributed throughout the entire cholesteric interface due to SDS and PVA surfactants. Those particles located near the newly-form disclination lines organize to form low energy configurations that stabilize both the particles and the defects (see below).  Meanwhile, particles in the neighboring homeotropic regions remain mobile at the interface and do not associate with defects---perhaps due to energy barriers associated with regions of partial planar alignment that flank both sides of the disclination lines (Fig.~\ref{SimulationsMain}B). These regions are captured only qualitatively by our simulations, which fail to describe the effects of adaptive anchoring present in experiment. The proposed assembly mechanism is supported by the FRAP measurements (Fig.~\ref{AnchtoDefect}E \& Supplementary Fig.~\ref{FRAPDiffSec}) and consistent with the large energy scales characterizing particle-defect interactions (of order $\xi K\gg k_B T$). 

To explain the observed pathway dependence of these particle-defect assemblies, we speculate that similar disclination lines formed in the absence of adsorbed particles migrate deeper into the liquid crystal bulk.  Such defects---like those shown in Fig.~\ref{AnchtoDefect}C---appear to be inaccessible to adsorbed particles introduced after their formation.  Thus, the kinetic processes of particle adsorption, osmotic swelling, and liquid crystal relaxation must act in concert to position particles in the right place at the right time to associate with disclination lines near the interface. 

\subsection*{Mapping out and stabilizing topological defects}

\begin{figure}
\centerline{\includegraphics[width=0.55\textwidth]{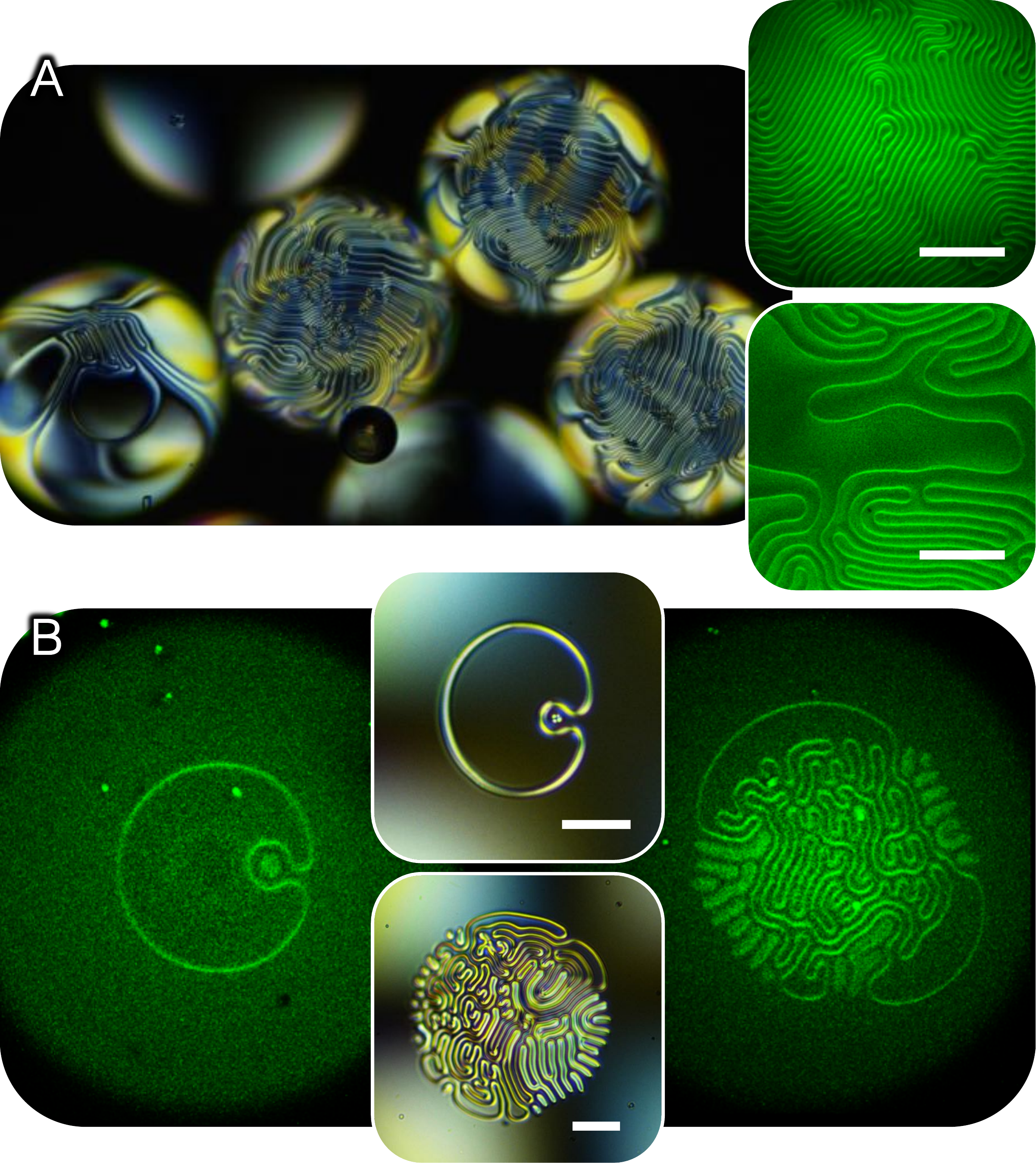}}
\caption{Particles aggregated along disclination lines can stabilize them, preventing them from annealing during osmotic swelling. After 4 days, shells swelled with excess colloids in the surrounding solution (A, from system in Supplementary Movie 1) have a higher density of disclinations that span across the shell, as compared to when excess nanoparticles are rinsed out (B, from system in Supplementary Movie 3). Skyrmions are stabilize in B) (seen in left panel), but not in A). Scale bars are 50 $\mu$m.\label{Skyrmion}}
\end{figure}

Not only can liquid crystal defects stabilize particle assembles, but the nanoparticles can stabilize defect structures that would otherwise vanish as the shells swell. In the absence of particles, swelling for 4 days with strong homeotropic anchoring causes the shells to become so thin that the anchoring energy overcomes the bulk elastic energy, completely untwisting the cholesteric and eliminating the surface stripes \cite{lt-prx}.  However, for shells left in the nanoparticle solution for this duration (Supplementary Movie 1), disclination lines remain and span across the entire shell as seen in polarized light and confocal micrographs (Fig.~\ref{Skyrmion}A). For shells in which excess particles are removed from the surrounding aqueous phase (Supplementary Movie 3), the area of defect lines is decreased (Fig.~\ref{Skyrmion}B). The correlation between nanoparticle concentration and defect area implies that the remaining defects are stabilized by the adsorbed particles. Interestingly, skyrmions appear and are mapped out by FluoSpheres only when excess nanoparticles are removed (Fig.~\ref{Skyrmion}B, \textit{left, top})  \cite{skyrmion-1}. The decreased number of adsorbed nanoparticles may allow disclination lines to anneal and become points of high twist, such as skyrmions. We believe these defects are skyrmions and not, for instance, torons because the stabilizing nanoparticles formerly followed planar anchoring, while torons do not have regions of planar anchoring surrounding them \cite{skyrmion-1,skyrmion-3,skyrmion-4}. Other types of cholesteric structures are also stabilized under conditions of reduced nanoparticle concentration, where cholesteric fingers form alongside disclinations (Fig.~\ref{Skyrmion}B, \textit{right, bottom})  \cite{oswald2000static, ivan-chfing}. In this way, the tendency of nanoparticles to become kinetically arrested within local energy minima can be harnessed for stabilizing liquid crystalline structures out of equilibrium.

\section*{Discussion}

The osmotic swelling of CLC shells alters the balance between bulk and surface energies, creating kinetic pathways by which adsorbed particles are repositioned from a surfactant-mediated state to a defect-mediated state. By balancing the rates of osmotic swelling and particle adsorption, the changing patterns of adsorbed nanoparticles both respond to and influence the cholesteric structures. The double emulsion platform we describe provides a means to control the time-varying energy landscape to guide the formation of particle-defect assemblies that depend on their kinetic history. Such linearly-aligned and kinetically-trapped particles act to stabilize the disclinations, preventing them from annealing. These results highlight opportunities for creating particle-liquid crystal assemblies of greater diversity by engineering the kinetic pathways by which they form. 

Realizing these opportunities will require a quantitative understanding of the competing elastic, defect, and surface contributions to the overall free energy. The Landau-de Gennes formalism provides a useful framework by which to describe these contributions; however, a systematic approach is needed to fully explore the space of relevant parameter regimes.  Cholesteric shells are characterized by multiple length scales: the shell radius and thickness, the pitch, the defect size, and the extrapolation length $K/W$ \cite{odl-klem}.  The addition of one or more particles to the interface introduces additional scales such as the particle diameter $d$, the interparticle spacing, as well as surface energies for the particle-liquid and particle-CLC interfaces\cite{iris}.  The present study focuses on one of many possible regimes in which the pitch is much smaller than the shell thickness but much larger than the particles; the extrapolation length is varied over a wide range spanning from pitch to the particle diameter to the defect size. We anticipate that changes in the relative magnitudes of these various scales will lead to qualitatively different energy landscapes, which can be used to guide formation of many other particle-CLC structures.

One significant feature absent from the current model is the adaptive anchoring condition at the CLC interface, whereby the absorption of particles and surfactants both influence and respond to the local liquid crystal alignment. The physico-chemical interactions between surfactant and liquid crystal molecules remain uncertain. In lieu of such molecular details, we advocate a phenomenological approach to describe how the surface energy depends on the concentration of adsorbed surfactants and the liquid crystal director. Boundary conditions incorporating these effects could be used to couple the liquid crystal bulk to the surfactant solution. Similar considerations could be applied to describe the distribution of ligands on the particle surface, which determines the relevant anchoring conditions on the particle. These conditions influence particle preference for planar or homeotropic regions at the surface and for bend or splay configurations in the bulk\cite{francesca}. The importance of particle size and ligand distribution in stabilizing certain states has already been demonstrated in other systems and should play a significant role here as well \cite{bishop-au}.

This study lays the groundwork for the fabrication of nanoparticle- and liquid crystal-based materials and for   fundamental investigations of interfacial self-assembly on liquid crystals. Photo-tunable cholesterics could be applied to create adaptable elastic landscapes through varying the pitch, potentially opening other kinetic avenues of assembly \cite{light-manipulate}. The pitch can also be decreased to imbue photonic properties to the system  \cite{photonic, beetle-mitov}. The ability to tune packing densities of particles at emulsion interfaces could also be applied to make colloidosomes with areas of differing permeabilities, potentially useful for encapsulation and controlled release applications  \cite{lee-osmswell,colloidosome}. Furthermore, for sensing applications, particles can be grafted with molecules that chemically respond to changes in their environment, whereby the underlying liquid crystal amplifies the appearance of reactions occurring at the particle surface  \cite{abblipid,sensing}. By considering the balance between bulk and surface energies, the particle assembly pathway can be engineered to shape colloidal assemblies,  topological defects, and their applications.

\section*{Methods}

\paragraph{Reagents, Sample Preparation, and Optical Characterization.}

The CLC is comprised of 5CB (4-cyano-4'-pentylbiphenyl, Kingston Chemicals Limited and Sython Chemicals) doped with CB15 ((S)-4-cyano-4-(2-methylbutyl)biphenyl, EMD Performance Materials and Sython Chemicals). Shells are produced using a microfluidic device with three nested, coaxial capillaries, as illustrated in  Fig.~\ref{ResultsIntro}B  \cite{ufluidics,lt-prx}. The water phases have 1\% wt polyvinyl alcohol (PVA, Sigma-Aldrich, 87-89\% hydrolyzed, average $M_{w} = 13-23$ kg/mol) to stabilize emulsions. Shells were subsequently stored within a solution of 1\% wt PVA, enclosed in a glass vial. 

To obtain ordered nanoparticle assemblies at the liquid crystal shell interface, a separate nanoparticle solution is first prepared with the corresponding, oppositely charged, hydrocarbon surfactant to render the particles partially hydrophobic (Fig.~\ref{ResultsIntro}C). Two types of fluorescent nanoparticles of different materials and sizes are used. 10-nm bare silica nanoparticles and 200-nm, amine-modified polystyrene nanoparticles (FluoSpheres) are purchased from Creative Diagnostics and Thermo Fisher Scientific, respectively. Both particles are introduced to the solution at a low concentration of 0.0001\% wt. Alkyl ammonium bromides (ATAB, Sigma-Aldrich) and sodium dodecylsulfate (SDS, Sigma-Aldrich) are used to modify silica and FluoSphere particles, respectively. Hydrochloric acid (HCl, Sigma-Aldrich) is added to the nanoparticle solution to adjust solution pH to be between 6 and 7, typically at a final concentration of 0.25 mM. Solutions are then sonicated for 30 minutes before gently mixing in liquid crystal shells and a concentrated PVA solution, to bring the final PVA concentration back to 1\% wt. We note that nanoparticle adsorption occurs more rapidly when PVA is added to the solution after the shells are introduced, with nanoparticle fluorescence signals at the interface noticeable after one hour. This is compared to signals seen after approximately a full day when PVA is sonicated with the initial nanoparticle-surfactant mixture. This implies that PVA interacts with the surfactant and the nanoparticles, slowing down interfacial particle adsorption.

Bright field and cross-polarized images are captured using two microscopes: a Nikon Eclipse Ti-U inverted microscope equipped with crossed polarizers and a Nikon DS-Ri2 camera, as well as an upright Zeiss AxioImager M2m microscope, also equipped with crossed polarizers and a Zeiss AxioCam HRc camera. Fluorescence micrographs are captured with both a Zeiss ApoTome attachment to the Zeiss AxioImage M2m microscope, to obtain both fluorescence and cross-polarized images of the same location, as well as a Zeiss LSM-700 laser scanning confocal microscope, to acquire higher resolution fluorescence images. Scanning laser wavelengths of 543 nm and 488 nm are used to excite the 10-nm silica and 200-nm FluoSphere nanoparticles, respectively. Fluorescence recovery after photobleaching (FRAP) measurements are performed by selecting regions for bleaching and gathering intensity measurements through the Zeiss Zen software. The time series of intensity measurements are subsequently analyzed using ImageJ.

\paragraph{Numerical Methods.}

We minimize a phenomenological Landau-de Gennes free energy fuctional  \cite{z-rav, lt-prx} to find the symmetric, traceless rank-2 tensor field $\mathbf{Q}(\mathbf{x})$ describing the nematic director. In the uniaxial limit, the director  $\mathbf{n}$ is related to $\mathbf{Q}$ by $\mathbf{Q}=(3S/2)\left[\mathbf{n} \otimes \mathbf{n}-\mathbf{I}/3\right]$, where $S$ is the nematic degree of order, $\otimes$ is a dyadic product (\textit{i.e.}, $[\mathbf{n} \otimes \mathbf{n}]_{ij} \equiv n_i n_j$) and $\mathbf{I}$ is the identity tensor.  The total free energy $\mathcal{F}$ includes volumetric contributions due to the bulk nematic order and spatial gradients thereof as well as surface contributions due to anchoring
\begin{equation}
    \mathcal{F} = \int_{\mathcal{V}} (f_{\mathrm{bulk}} + f_{\mathrm{grad}})  \mathrm{d}V + \int_{\mathcal{S}} f_{\mathrm{surf}} \mathrm{d}S \label{eq:F}.
\end{equation}
The bulk contribution to the free energy density is given by
\begin{equation}
f_{\mathrm{bulk}}= \frac{A}{2}  \Tr \mathbf{Q}^2  + \frac{B}{3} \Tr \mathbf{Q}^3+ \frac{C}{4}( \Tr \mathbf{Q}^2)^2,
\end{equation}
where $\Tr$ is the trace. A uniaxial $\mathbf{Q}$-tensor minimizes this free energy with a value of $S=S_0\equiv(-B+\sqrt{B^2-24AC})/6C$. The  gradient contribution to the free energy density is
\begin{equation}
f_{\mathrm{grad}}= \frac{L}{2} (\nabla \times \mathbf{Q}+2q_0 \mathbf{Q})^2 + \frac{L}{2} (\nabla \cdot \mathbf{Q})^2,
\end{equation}
where $L$ is the elastic constant, and $2\pi/q_0$ is the cholesteric pitch. This approximate free energy density uses a single elastic constant and neglects additional contributions from saddle-splay distortions.  Finally, the free energy (surface) density due to anchoring is modeled using the Rapini-Papoular surface potential. Homeotropic anchoring at surface $\mathcal{S}_0$ is described by the free energy density
\begin{equation}
f_{\mathrm{surf}} = W_0 \Tr [(\mathbf{Q}-\mathbf{Q}^{\parallel})^2]\quad \text{for } \mathbf{x}\in\mathcal{S}_0,
\end{equation}
where $W_0$ is the homeotropic anchoring strength, and $\mathbf{Q}^{\parallel}= (3 S_0/2)( \bm{\nu} \otimes \bm{\nu}-\mathbf{I}/3)$ is the uniaxial $\mathbf{Q}$-tensor oriented parallel to the surface normal vector $\bm{\nu}$. Planar anchoring at surface $\mathcal{S}_1$ is described by the free energy density 
\begin{equation}
f_{\mathrm{surf}} = W_1 \left[ \Tr[(\bar{\mathbf{Q}}-\bar{\mathbf{Q}}^{\perp})^2] +(\Tr \mathbf{Q}^2-3 S_0^2/2)^2 \right] \quad \text{for } \mathbf{x}\in\mathcal{S}_1,
\end{equation}
where $W_1$ is the planar anchoring strength, $\bar{\mathbf{Q}} \equiv \mathbf{Q}+S_0 \mathbf{I}/2$,  and $\bar{\mathbf{Q}}^{\perp}=(\mathbf{I}- \bm{\nu} \otimes \bm{\nu}) \bar{\bm{Q}} (\mathbf{I}- \bm{\nu} \otimes \bm{\nu})$ is a  projection of the $\mathbf{Q}$ tensor on the plane perpendicular to the surface normal $\bm{\nu}$. 

We use the following parameter estimates corresponding to the values for 5CB given in Ref.\ \cite{z-rav}: $A=-0.172\times 10^6$ J/m$^3$, $B=-2.12\times10^6$ J/m$^3$, $C=1.73\times10^6$ J/m$^3$,  $L=2\times10^{-11}$ N; the pitch is specified to be $2\pi/q_0=0.25~\mu$m. With these parameters, the nematic degree of order at equilibrium is $S_0=0.533$, and the nematic correlation length is $\xi =6.63$ nm, which sets the characteristic size of defects. For the one-constant approximation used here, the Frank elastic constant is $K=9L S_0^2/2 = 2\times10^{-11}$ N.

The total free energy (\ref{eq:F}) is minimized using the conjugate gradient method from the ALGLIB package (http://www.alglib.net/) to determine the equilibrium tensor field. The three-dimensional rectangular domain is discretized using a cubic mesh of size $\Delta x=4.4$ nm. For each candidate $\mathbf{Q}(\mathbf{x})$, the required gradients and integrals are approximated using finite difference energy minimization to compute the total free energy. For the simulations presented here, the $\mathbf{Q}(\mathbf{x})$ tensor field is initialized as the cholesteric ground state with the pitch axis oriented along the $\hat{z}$ direction. We note that the final configuration corresponds to a local minimum of the energy functional (not necessarily the global minimum).

\section*{Data availability}
The data that support the findings of this study are available from the corresponding author upon reasonable request.


\bibliography{manuscript} 
\bibliographystyle{ieeetr}

\section*{Acknowledgements}

We thank Guillaume Durey, Martin F. Haase, Randall D. Kamien, Maxim O. Lavrentovich, Teresa Lopez-Leon, Kathleen J. Stebe, and Slobodan \v{Z}umer for helpful discussions. L.T. acknowledges support from the Simons Society of Fellows of the Simons Foundation.

\section*{Author Contributions}

L.T. designed research. L.T. performed research. L.T. and K.J.M.B analyzed data. L.T. and K.J.M.B. wrote the paper.

\section*{Competing Interests}

The authors declare no competing interests.


\newpage

\beginsupplement

\clearpage

\begin{figure*}[bt!]
\section*{Supporting Information}

\noindent \textbf{Movie S1.} Cholesteric liquid crystal double emulsions osmotically swell in an aqueous phase with 1\% wt polyvinyl alcohol (PVA), 10 mM sodium dodecylsulfate (SDS), 0.25 mM hydrochloric acid (HCl), and 0.0001\% wt FluoSphere. The inner aqueous phase has 1\% wt PVA and 0.5 M sucrose. One $z$-scan is taken every 10 minutes, and is projected onto the $xy$-plane for each frame. The rapid evolution of shell patterning towards subsurface defect lines correlates with increase in fluorescence intensity at the shell surface. This implies that as more particles deposit onto the cholesteric interface, the homeotropic anchoring strength increases. 
\\

\noindent \noindent \textbf{Movie S2.} Cholesteric liquid crystal double emulsions are equilibrated in a solution of 1\% wt PVA, 7 mM SDS, 0.25 mM HCl, and 0.5 M sucrose. This results in shells with weak homeotropic anchoring as an initial condition. Shells are then transferred to a solution with 1\% wt PVA, 10 mM SDS, 0.25 mM HCl, and 0.0001\% wt FluoSpheres to initiate osmotic swelling and particle deposition. Fluorescence intensities across the region marked by the white line (right) are plotted on the left for each frame of the time series (right). Peaks in the fluorescence intensity follow disclination lines as the pattern evolves with time, indicating particles locating along defects. As homeotropic anchoring increases, defect lines separate in order to accommodate more regions of perpendicular alignment between them.
\\

\noindent \textbf{Movie S3.} Cholesteric liquid crystal double emulsions osmotically swell in an aqueous phase with 1\% wt polyvinyl alcohol (PVA), 10 mM sodium dodecylsulfate (SDS), 0.25 mM hydrochloric acid (HCl), and 0.0001\% wt FluoSphere. The inner aqueous phase has 1\% wt PVA and 0.5 M sucrose. After several minutes, emulsions are removed from the nanoparticle solution and then pipetted into a solution with 1\% wt PVA, 10 mM SDS, 0.25 mM HCl, and without particles. This is done to remove excess particles surrounding the shells and to reduce the amount of particles that interfacially adsorb to the cholesteric interface. This steadies the fluorescence intensities across the time series, allowing for observation of the repositioning of FluoSpheres from homeotropic anchoring regions to defects. Analysis of fluorescence intensity profiles of stripes in the time series is plotted in Fig.3D.
\\
\end{figure*}

\clearpage

\clearpage

\end{document}